\documentclass[draft]{agujournal}

\journalname{JGR-Planets}

\begin{document}

%
%

\title{Constraining the potential liquid water environment at Gale crater, Mars}

%
%

\authors{Edgard G. Rivera-Valent\'{i}n\affil{1,2}, Raina V. Gough\affil{3,4}, Vincent F. Chevrier\affil{5}, Katherine M. Primm\affil{3,4}, German M. Mart\'{i}nez\affil{6}, Margaret Tolbert\affil{3,4}}

\affiliation{1}{Arecibo Observatory, Universities Space Research Association, Arecibo, PR 00612, USA}
\affiliation{2}{Lunar and Planetary Institute, Universities Space Research Association, Houston, TX 77058, USA}
\affiliation{3}{Cooperative Institute for Research in Environmental Sciences, University of Colorado, Boulder, CO 80309, USA}
\affiliation{4}{Department of Chemistry and Biochemistry, University of Colorado, Boulder, CO 80309, USA}
\affiliation{5}{Arkansas Center for Space and Planetary Sciences, University of Arkansas, Fayetteville, AR 72701, USA}
\affiliation{6}{Department of Climate and Space Sciences and Engineering, University of Michigan, Ann Arbor, MI, USA}


\correspondingauthor{Edgard G. Rivera-Valent\'{i}n}{ervalentin@usra.edu}


\begin{keypoints}
\item Measured surface environmental conditions at Gale crater, Mars are not favorable to brine formation via deliquescence of calcium perchlorate.  
\item Liquids may have formed in the shallow subsurface of low thermal inertia units within MSL-traversed terrains. 
\item In terms of operation strategies, MSL may best find liquids in the subsurface of units with thermal inertia $\lesssim175$~J~m$^{-2}$~K$^{-1}$~s$^{-1/2}$ and albedo $\gtrsim0.25$ around Ls 100$^{\circ}$.  
\end{keypoints}

%
%

\begin{abstract}
The Mars Science Laboratory (MSL) Rover Environmental Monitoring Station (REMS) has now made continuous in-situ meteorological measurements for several martian years at Gale crater, Mars. Of importance in the search for liquid formation are REMS' measurements of ground temperature and in-air measurements of temperature and relative humidity, which is with respect to ice. Such data can constrain the surface and subsurface stability of brines. Here we use updated calibrations to REMS data and consistent relative humidity comparisons (\emph{i.e.,} w.r.t. liquid vs w.r.t. ice) to investigate the potential formation of surface and subsurface liquids throughout MSL's traverse. We specifically study the potential for the deliquescence of calcium perchlorate. Our data analysis suggests that surface brine formation is not favored within the first 1648 sols as there are only two times (sols 1232 and 1311) when humidity-temperature conditions were within error consistent with a liquid phase. On the other hand, modeling of the subsurface environment would support brine production in the shallow subsurface. Indeed, we find that the shallow subsurface for terrains with low thermal inertia ($\Gamma\lesssim 300$~J~m$^{-2}$~K$^{-1}$~s$^{-1/2}$) may be occasionally favorable to brine formation through deliquescence. Terrains with $\Gamma\lesssim175$~J~m$^{-2}$~K$^{-1}$~s$^{-1/2}$ and albedos of $\gtrsim0.25$ are the most apt to subsurface brine formation. Should brines form, they would occur around Ls 100$^{\circ}$. Their predicted properties would not meet the Special nor Uncertain Region requirements, as such they would not be potential habitable environments to life as we know it. 

\textbf{Plain Language Abstract}

The Mars Science Laboratory (MSL) has now made continuous measurements of the local weather at Gale crater, Mars. Such measurements can help guide our search for the formation of liquid water on present-day Mars. Specifically, when the right temperature and humidity conditions are met, certain salts can take in water vapor from the atmosphere to produce liquids. Here, we use data from MSL along with experimental results on the stability of a Mars-relevant salt to search for time periods when liquids could potentially form at the surface. Additionally, we use simulations and MSL data to understand the potential to form such liquids in the subsurface. Our results suggest that surface formation of liquids are unlikely throughout MSL's travels; however, the shallow subsurface may experience conditions that would allow for liquid formation. However, not much liquid would form, and the properties of these liquids would not permit life as we know it to persist.

\end{abstract}

%
%

\section{Introduction}
At large spatial scales, the present martian atmospheric conditions preclude the formation of pure liquid water at the surface \citep{Haberle:2001}. The formation of transient brines, though, has long been hypothesized since the discovery that the martian surface is composed of a few wt\% salt \citep{Ingersoll:1970, Clark:1978, Brass:1980}. Perchlorate (ClO$_{4}^{-}$) salts have now been identified in-situ by the Phoenix lander \citep{Hecht:2009}, potentially at the two Viking sites \citep{NavarroGonzalez:2010}, by the Mars Science Laboratory (MSL) Curiosity rover \citep{Leshin:2013, Ming:2014}, possibly in the form of hydrated calcium perchlorate (Ca(ClO$_{4}$)$_{2}\cdot$nH$_{2}$O) \citep{Glavin:2013}, and from orbit at sites with recurring slope lineae (RSL) activity \citep{Ojha:2015}. Such salts are interesting because of their ability to transition from a solid crystalline salt into an aqueous solution given the appropriate temperature and relative humidity (\emph{i.e.,} deliquescence) \citep{Zorzano:2009, Gough:2011, Gough:2014, Fischer:2014, Nuding:2014, Nikolakakos:2017}. Calcium perchlorate can produce brines above temperatures of 198~K \citep{Pestova:2005, Marion:2010}, the lowest known eutectic temperature for pure component brines of Mars-relevant salts. Calcium perchlorate has been shown to deliquesce at a relative humidity with respect to liquid of $RH_{l}\sim$50\% and not effloresce (\emph{i.e.,} transition from aqueous to solid) until a much lower $RH_{l}\sim$3\% \citep{Nuding:2014}, making it ideal for liquid production on present-day Mars. Furthermore, liquid formation through deliquescence of calcium perchlorate has been shown to allow for a viable environment for microorganisms under Mars-like conditions \citep{Nuding:2017}, suggesting a potentially astrobiologically interesting process. 

Indeed, the search for liquid water on present-day Mars has largely been driven by astrobiology, but also as a potential trigger mechanism or fluid for mass wasting events, such as gullies \citep{Malin:2003, Johnsson:2014, Masse:2016}, slope streaks \citep{Sullivan:2001, Kreslavsky:2009}, dark dune spots \citep{Kereszturi:2010, Kereszturi:2012}, and RSL \citep{McEwen:2011,  McEwen:2013, Chevrier:2012, Dundas:2017}. Potential visible confirmation of liquid water, though, has only come in the form of droplets on the lander legs of Phoenix \citep{Renno:2009}. At Phoenix, the presence of liquid water was also inferred by the heterogeneous distribution of salts within the regolith \citep{Cull:2010}, which could potentially have been relocated by thin films of solutions, and by the measured changes in regolith dielectric signatures during nighttime \citep{Stillman:2011}. Such interpretations of liquids at the Phoenix landing site have further been supported by experimental results that suggest a role for deliquescence \citep{Chevrier:2009, Nuding:2014, Fischer:2016}. 

Liquid production through deliquescence, though, has been suggested to be constrained to a select few regions on Mars \citep{Martinez:2013, Kossacki:2014}. However, recently, surface meteorological conditions derived from the Rover Environmental Monitoring Station (REMS) suggested favorable conditions for brine production through deliquescence of calcium perchlorate at the surface and shallow subsurface of Gale crater \citep{MartinTorres:2015}. REMS data, though, has undergone several calibrations \citep{Martinez:2017}, the last of which was on June 2015. The latest calibrations have lead to drier conditions, but, more importantly, previous investigations have compared relative humidity w.r.t. ice ($RH_{i}$) as measured by REMS to a phase diagram w.r.t. liquid water (\emph{e.g.,} \citet{Rummel:2014, MartinTorres:2015, Gough:2016, Martinez:2017, Pal:2017}). Indeed, MSL REMS measures air relative humidity at a height of 1.6~m using a capacitance based hygrometer calibrated to $RH_{i}$ \citep{Harri:2014}. Under the low temperatures measured at Gale crater ($\sim$180~K) \citep{Hamilton:2014}, the difference between $RH_{l}$ and $RH_{i}$ can be on the order of 50\% because of the significant difference between the respective saturation vapor pressures. This difference could drastically alter interpretation of potential brine formation events through deliquescence. Here we use consistent relative humidity comparisons when analyzing MSL REMS temperature and relative humidity data in search of favorable conditions on the surface for brine formation via deliquescence of calcium perchlorate. Additionally, we simulate the subsurface environment at Gale crater to investigate the potential for deliquescence of calcium perchlorate in order to provide guidance to future operation strategies for the rover and elucidate the role of deliquescence in the near-surface martian water cycle at equatorial regions. 

%
%

\section{Data Analysis}
The Mars Science Laboratory (MSL) Curiosity rover landed on the floor of Gale crater (4.7$^{\circ}$S, 137$^{\circ}$E) on August 5, 2012 and, as of January 2018, has been operating for more than 1900 sols (almost 3 Martian years). Among its ten science instruments \citep{Grotzinger:2012}, the REMS is a suite of sensors designed to assess the environmental conditions along Curiosity's traverse \citep{GomezElvira:2012, Hamilton:2014, Harri:2014}. The REMS instrument includes six sensors that measure ground and air temperature, wind velocity and relative humidity w.r.t. to ice at 1.6~m height on the rover mast, and atmospheric pressure and UV radiation at about 1~m height on the rover deck. The REMS nominal strategy for data acquisition consists of five minutes of measurements at 1~Hz every Mars hour, with interspersed full hour sample periods at 1~Hz to cover every time of the sol over a period of a few sols. 
s
In this work, we focus on REMS measurements performed by the relative humidity sensor (RHS) and ground temperature sensor (GTS) during the first 1648 sols. The RHS consists of three polymeric sensors that measures the air relative humidity with respect to ice ($RH_{i_{a}}$) and a sensor that measures the temperature of the air ($T_{a}$) entering the RHS \citep{Harri:2014}. The uncertainty in $RH_{i_{a}}$ is $\pm20$\% for $T_{a}<203$~K, $\pm10$\% for 203~K $\leq T_{a}\leq$ 243~K, and $\pm2$\% for $T_{a} > 243$~K, while for $T_{a}$ the uncertainty is 0.2 K. Among the full set of RHS measurements, we only consider those taken during the first four seconds after the RHS has been turned on after at least $\sim$5~min of inactivity. After four seconds, the RHS is affected by heating produced by the sensor itself, which increases its temperature by up to 1.5~K. Thus, the local $RH_{i_{a}}$ measured by the sensor is closest to the actual value of the atmosphere only during the first few seconds of operation. Reliable $RH_{i_{a}}$ values include measurements taken during the nominal and the so-called high resolution interval mode (HRIM), which consists of alternately switching the sensor on and off at periodic intervals to minimize heating and is only used on selected one to two hour observation blocks (see Fig. 13 in \citet{Martinez:2017}). Since during these four seconds $RH_{i_{a}}$ values stay stable, this strategy typically results in 24 hourly values of $RH_{i_{a}}$ and $T_{a}$ under nominal measuring strategy.

The GTS measures the intensity of infrared radiation emitted by the ground in the bandwidths 8 - 14, 15.5 - 19, and 14.5 - 15.5 $\mu$m, from which the surface brightness can be derived \citep{Sebastian:2010}. It uses three thermopiles pointed 26$^{\circ}$ downward from the plane of the rover deck with a field of view of 60$^{\circ}$ horizontally and 40$^{\circ}$ vertically, covering a ground area of about 100~m$^{2}$ depending on the angle between the surface and sensor. For details on random and systematic uncertainties on GTS measurements, see \citet{Hamilton:2014} and the supplementary material in \citet{Martinez:2016}. Among the full set of GTS measurements, we only consider here those with the highest confidence possible. Specifically, those with the ASIC power supply in range, the highest recalibration quality, and with no shadows in the GTS field of view. By averaging these GTS measurements over 5-min-long intervals centered at simultaneous RHS measurements of the highest confidence, and by imposing that at least 60 GTS measurements must be averaged to reduce noise, we produce high quality hourly values of ground temperature ($T_{g}$) simultaneous to $RH_{i_{a}}$.

Using simultaneous hourly values of the highest possible confidence of $RH_{i_{a}}$ and $T_{g}$, ground relative humidity was inferred assuming water vapor pressure is constant throughout the 1.6~m air column such that $RH_{i_g} = RH_{i_a}\left[p_{sat_i}(T_{a})/p_{sat_i}(T_{g})\right]$, where $p_{sat_i}$ is the saturation vapor pressure for ice \citep{Feistel:2007}. Water vapor pressure, which can be inferred by $P_{H_{2}O}=RH_{i_{a}}\times p{sat_i}\left(T_{a}\right)$, is not expected to be constant from ground to a height of 1.6~m. Indeed, the gradient between ground and air will vary throughout the day as the planetary boundary layer (PBL) height changes \citep{Zent:1993, Savijarvi:2016}. During daytime there does not exist a strong gradient between surface and a height of 1.6~m because the PBL height approaches the scale height allowing for vigorous mixing of water vapor; however, at nighttime, the PBL is thin and gradients in water vapor between near-surface and air form \citep{Zent:1993}. In the absence of a reliable way to extract water vapor pressure at the ground, though, we make this simplifying assumption. 

Propagating error from REMS measurements, we found that most low $RH_{i_g}$ have large errors that do not preclude the unphysical possibility of $RH_{i_g}<0\%$; therefore, these values were not included in our analysis. This filter on the data results in a loss of information during the warmest periods at Gale crater, which approached temperatures of $\sim290$~K as measured by GTS. Additionally, some REMS measurements suggest $RH_{i_g}>>100\%$, even up to 500\%, which is unrealistic. We note, though, that super saturation w.r.t. ice has been shown to occur at high altitudes within the martian atmosphere \citep{Maltagliati:2011} and ice supersaturation of $RH_{i}\sim170$\% are often observed in Earth's atmosphere \citep{Jensen:2013}.  For completeness, we present data in both $RH_{l}$ and $RH_{i}$ space up to saturation within the relevant phase space. 

\subsection{Surface deliquescence of calcium perchlorate}
Typically, the phase diagram of brines is presented with respect to temperature and water activity ($a_{w}$) or concentration in wt\%, which also relates to $a_{w}$, (e.g., \citet{Chevrier:2009, Gough:2011, Hanley:2012, Gough:2014, Nuding:2014}). Under equilibrium conditions, $a_{w}=RH_{l}/100$ \citep{Murphy:2005}, and so \emph{in-situ} humidity data can be directly compared with the phase diagram of salts after some modification. Fundamentally, though, a liquid forms when the partial pressure of water is above the partial pressure at the eutectic point and simultaneously the temperature is above the eutectic temperature; therefore, the most direct way to search for liquid production would be to compare \emph{in-situ} inferred $P_{H_{2}O}$ and temperature with a phase diagram in the same terms. However, thus far \emph{in-situ} humidity measurements on Mars have been measurements of $RH_{i}$. Inferring $P_{H_{2}O}$ from $RH_{i}$ and temperature, and propagating the errors in both measurements, typically leads to large errors in $P_{H_{2}O}$ because of the exponential dependency in $p_{sat_i}$. Here, in order to search for environmental conditions appropriate for the deliquescence of calcium perchlorate, we compared REMS data to the phase diagram of Ca(ClO$_{4}$)$_{2}$ in $RH_{l}$ (Figure 1a) and in $RH_{i}$ phase space (Figure 1b) to be consistent with previous results and to more confidently search for humidity-temperature conditions that could permit liquid formation. We also compare with respect to $P_{H_{2}O}$ (Figure 1c) for completeness. 

REMS inferred $RH_{i_g}$ was converted to relative humidity w.r.t. liquid by $RH_{l_g}=RH_{i_a}\left[p_{sat_i}(T_{a})/p_{sat_l}(T_{g})\right]$, where $p_{sat_l}$ is the saturation vapor pressure of liquid water, equation 10 from \citet{Murphy:2005}. Modeled deliquescence relative humidity (DRH) is presented in blue in Figure 1 \citep{Nuding:2014}. Thermodynamically, efflorescence relative humidity (ERH) is indistinguishable from DRH; however, experimental work has shown that calcium perchlorate, like many salts, undergoes a hysteresis effect, efflorescing at much lower relative humidity conditions. Here we use fits to experimental data from \citet{Nuding:2014} to extrapolate the ERH behavior of Ca(ClO$_{4}$)$_{2}$ as a function of temperature (red in Figure 1). In Figure 1a, the ice line (\emph{i.e.,} $RH_{i}=100\%$ in $RH_{l}$ phase space) is presented as a black solid line. 

%
\begin{figure}[h!]
\includegraphics[width=18pc]{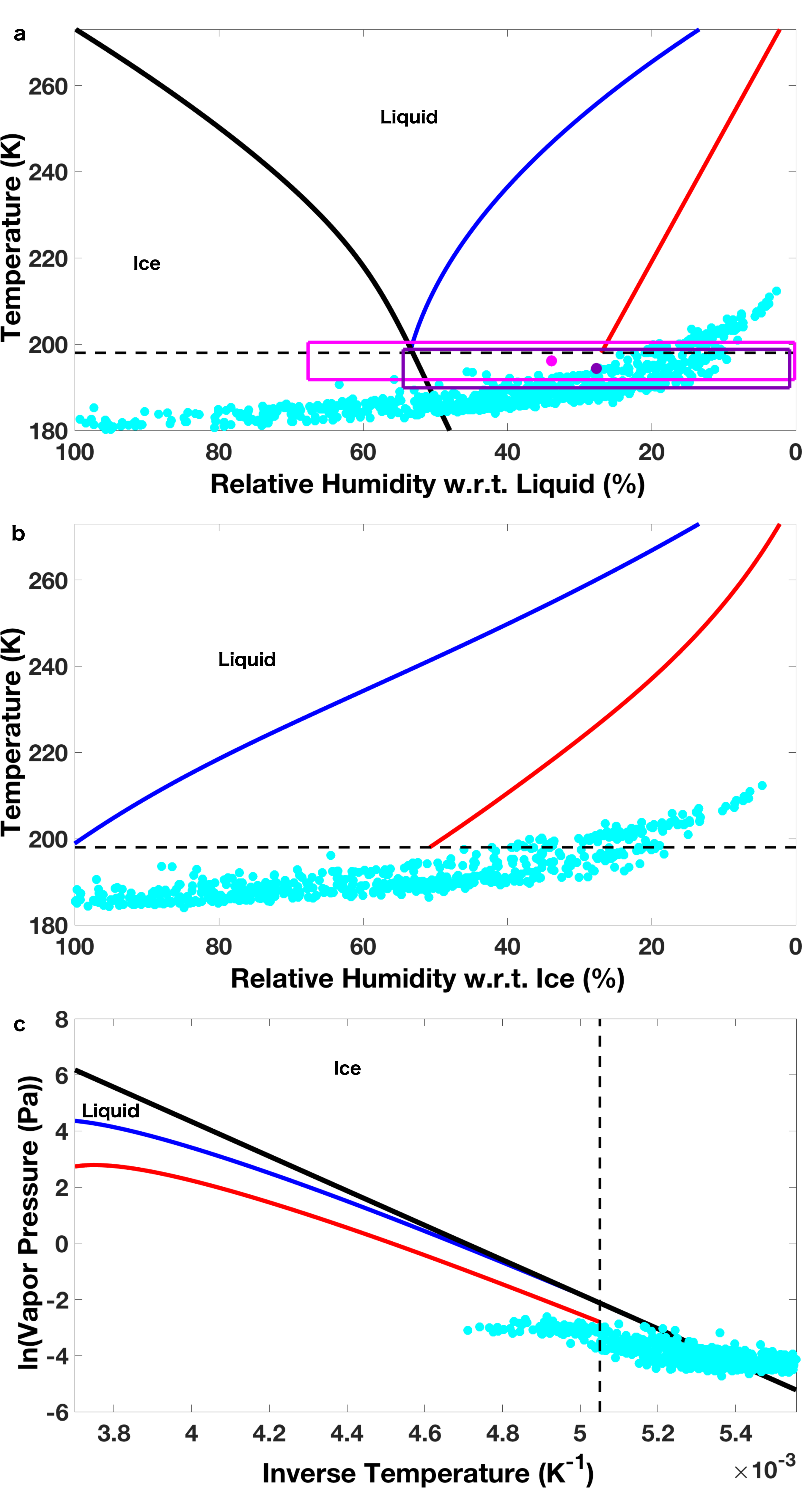}
\caption{MSL REMS measured ground temperature and inferred ground relative humidity on the phase diagram of calcium perchlorate in relative humidity (a) w.r.t. liquid, (b) w.r.t. ice phase space, and (c) in 1/$T_{g}$ vs ln($P_{H_{2}O}$) space. The blue and red lines are the deliquescence relative humidity and efflorescence relative humidity respectively from \citep{Nuding:2014} while REMS data is in cyan circles. The black dashed line is the eutectic temperature of calcium perchlorate brine, 198~K. In (a) the solid black line is the ice line, where $RH_{i}=100\%$ in $RH_{l}$ phase space, while in (b) it is at the left-hand plot axis, and in (c) the black solid line is the saturation vapor pressure of ice as a function of temperature. In (a) measurements for sol 1232 (magenta) and 1311 (purple) are shown with 2-sigma error in temperature and 1-sigma in relative humidity, illustrated by the respectively colored box. Because error in relative humidity is large, even at the 1-sigma, we do not investigate the 2-sigma effect. }
\end{figure}

Accounting for updated REMS data calibrations and comparing relative humidity values in consistent phase space, we found that no measured environmental condition would permit for the deliquescence of calcium perchlorate at the surface. Accounting for error, we found that to the 1-sigma level for $T_{g}$ and $RH_{l_g}$ deliquescence is still unfavored; however, to the 2-sigma level for $T_{g}$, 1-sigma for $RH_{l_g}$, there are two points on sols 1232 and 1311, Ls 99$^{\circ}$ and Ls 137$^{\circ}$ respectively, that could be within the liquid phase. These values, delineated in Figure 1a, occurred while the rover was near active sand dunes \citep{Vasavada:2017} during the early morning and late evening. Accounting for 2-sigma level error in $T_{g}$ would still preclude the formation of liquids via deliquescence of other hygroscopic salts, such as magnesium perchlorate, the next most stable Mars-relevant single component brine, which has a eutectic temperature of 205~K and eutectic concentration of $a_{w}=0.55$ \citep{Gough:2011}.

Our conclusion, that surface humidity-temperature conditions are not favorable to liquid production, is valid only for single-salt brines. Multi-component brines, which would be expected on Mars, would have lower DRH compared to each individual salt. Such multi-component solutions would also have lower eutectic temperatures. Indeed, the DRH of binary mixtures was found to be lower than that of the least deliquescent salt in the system \citep{Gough:2014}. Therefore, multi-component brines may be more stable on the martian surface. 

\subsection{Surface enthalpy changes}
Active near-surface processes, such as sublimation, condensation, hydration state changes, and deliquescence, result in enthalpy changes ($\Delta H$) that can be inferred from water vapor pressure ($P_{H_{2}O}$) and $T_{g}$, provided equilibrium conditions are assumed (\emph{e.g.,} \citet{Zent:2010, RiveraValentin:2015}). The inferred $\Delta H$ is the sum of all active processes during the studied timespan and so this procedure can provide information on the dominant ongoing near-surface processes. Thus, following the methods of \citet{RiveraValentin:2015}, we searched for surface enthalpic changes during nighttime as a potential signature of phase changes, specifically within 3 hour timespans from midnight to 3am, 3am to 6am, 6am to 9am, 6pm to 9pm, and 9pm to midnight. 

Vapor pressure curves were constructed using the REMS measured surface temperatures along with the inferred water vapor pressure at 1.6~m. Data was binned over five sols for each studied timespan to increase the statistics. Following the Clausius-Clapeyron relation for transitions between gas and a condensed phase, where 
\begin{equation}
\ln\left(P_{H_{2}O}\right)=-\frac{\Delta H}{R}\left(\frac{1}{T_{g}}\right)+c,
\end{equation}
such that the slope in 1/$T_{g}$ vs ln($P_{H_{2}O}$) space is linear and related to enthalpy by $\Delta H = -\beta R$ \citep{Murphy:2005, RiveraValentin:2015}, where $\beta$ is the slope, $R$ is the ideal gas constant, and $c$ is a constant. Here, we used a weighted least squares fit method to derive $\beta$. The weighted least squares fit slope, which accounts for error following the York method \citep{York:2004} of each MSL measurement in $T_{g}$ and inferred $P_{H_{2}O}$, was found to 90\% confidence. To test for statistical significance of $\beta$, we test for the null hypothesis (\emph{i.e.,} that  1/$T_{g}$ contributes no information for the prediction of ln($P_{H_{2}O}$)); therefore, only non-zero slopes within error, and so non-zero $\Delta H$, are accepted.  

We found only nine significant (\emph{i.e.,} non-zero) enthalpic changes, all of which occurred during the early morning (six values during 3am to 6am, and three values from 6am to 9am) primarily around Ls 100$^{\circ}$. Inferred values with associated error, including propagation from measurement error, are shown in Figure 2, where colors delineate time span, as a function of Ls. Cumulatively, the derived values have a weighted average of $\Delta H = 33 \pm 20$~kJ/mol. Derived enthalpic changes were generally within error of the enthalpy of H$_{2}$O sublimation (50.9~kJ/mol), but the range of values does not preclude other processes such as adsorption/desorption, which has been suggested to be active within Gale crater \citep{Savijarvi:2016}, or deliquescence \citep{Jia:2018}. Of note, a non-zero derived enthalpy on Ls 99$^{\circ}$ (sol 1232) could support liquid formation as indicated from the analysis in section 2.1 for sol 1232 after including 2-sigma error; however, no corresponding statistically significant $\Delta H$ was found for sol 1311. Derived values, though, agree with the hour (4am to 6am) and Ls range during which frost formation was likely within Gale crater \citep{Martinez:2016}; therefore, derived enthalpic changes may support frost/sublimation as an active water vapor sink/source within Gale crater.  

%
\begin{figure}[h]
\includegraphics[width=30pc]{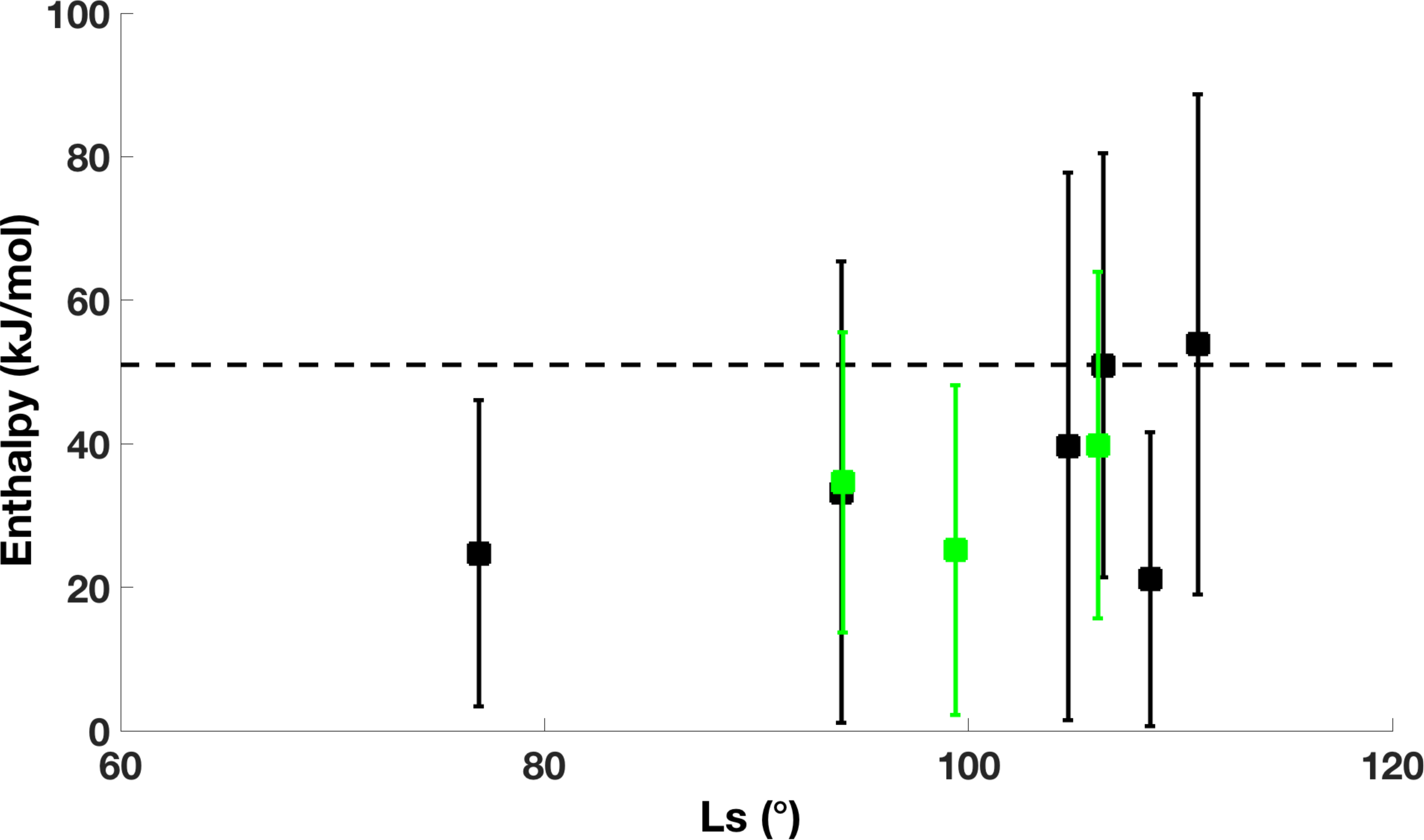}
\caption{Derived non-zero enthalpies over 5 sols from timespans of 3am to 6am (black squares) and 6am to 9am (green squares) with errorbars to 90\% confidence as a function of Ls. The black dashed line is the enthalpy of sublimation of water ice (50.9~kJ/mol).}
\end{figure}

%
%

\section{Subsurface Modeling}
MSL has traversed various terrain types, which have included both low and high thermal inertia terrains that have deviated from the typical values found at the landing ellipse ($\Gamma\sim350$~J~m$^{-2}$~K$^{-1}$~s$^{-1/2}$) \citep{Christensen:2001, Putzig:2005}. Reported thermal inertia values in the first 1337 sols have ranged from 170~J~m$^{-2}$~K$^{-1}$~s$^{-1/2}$ to 600~J~m$^{-2}$~K$^{-1}$~s$^{-1/2}$ and albedo values have typically been $0.1\leq A\leq0.3$ \citep{Martinez:2014, Martinez:2016, RodriguezColon:2016, Vasavada:2017}. Via a fully coupled heat and mass transfer model, we searched for potential subsurface liquid formation throughout one martian year at locations the rover has traversed. Additionally, we explored liquid formation for various combinations of regolith thermal properties in order to better inform future MSL operational strategies and elucidate the potential for subsurface brine formation at equatorial regions on Mars using Gale crater as a proxy. 

\subsection{Methods}
Subsurface temperatures were simulated by solving the 1-D thermal diffusion equation via a finite element procedure \citep{RiveraValentin:2011, Chevrier:2012, Kereszturi:2012, Nuding:2014} with a vertical resolution of 0.01~m. The time step required for stable solutions is dependent on the thermal inertia of the regolith column and the vertical resolution; values used here ranged from 180~s to 370~s. The surface boundary condition is radiative and includes direct illumination, along with scattering and thermal emission atmospheric components such that the incident heat flux is given by
\begin{equation}
Q_{i}=\left(1-A\right)\frac{S_{0}}{r^{2}}\left[\psi\cos\left(\zeta\right) +\left(1-\psi\right)f_{scat}+\epsilon\cos\left(\theta-\phi\right)f_{atm}\right] ,
\end{equation}
where $S_{0}$ is the solar flux at 1 AU, $r$ is the instantaneous sun-Mars distance in AU, $\zeta$ is the solar zenith angle, $\psi$ is the transmission coefficient, $f_{scat}$ and $f_{atm}$ are the fractional amounts of the relevant flux reaching the surface, $\epsilon$ is atmospheric thermal emissivity, and $\theta$ and $\phi$ are the solar declination and latitude respectively \citep{Applebaum:1989, Pollack:1990p2, Aharonson:2006, Schmidt:2009}. As applied by \citet{Blackburn:2009}, $\psi$ is a polynomial fit to data from \citet{Pollack:1990p2} as presented in \citet{Rapp:2008}. By \citet{Schmidt:2009}, $f_{scat}=0.02$, $f_{atm}=0.04$, and $\epsilon=0.9$. A flat surface is assumed and so slope effects on thermal insolation were not accounted for (\emph{e.g.,} \citet{Aharonson:2006}). Note, such simulations provide spatially and temporally averaged temperatures. Actual values may vary due to differences in regolith physical properties with depth and local geometry; however, the code has been validated against PHX \citep{RiveraValentin:2015} and MSL \citep{RodriguezColon:2016} measurements, and was further validated here. 

Water vapor diffusion through regolith, which has been shown to be approximately Fickian \citep{Clifford:1986} and undergo diffusion advection \citep{Ulrich:2009}, follows 
\begin{equation}
J_{DA}=\frac{\varphi}{\tau \mu}\left(D_{H_{2}O/CO_{2}}\frac{P}{RT}\frac{d\gamma}{dz}+J_{DA}\right),
\end{equation}
where $D_{H_{2}O/CO_{2}}$ is the diffusivity of water vapor through CO$_{2}$ gas, $\varphi=0.5$ \citep{Zent:2010}, and $\tau=2$ \citep{Hudson:2008, Sizemore:2008} are the porosity and tortuosity respectively, $\mu$ is the ratio between the molecular weights of H$_{2}$O and CO$_{2}$, $P$ is air pressure, and $\gamma$ is the water vapor mixing ratio. The diffusivity of water vapor through CO$_{2}$ gas was modeled as temperature dependent following
\begin{equation}
D_{H_{2}O/CO_{2}} = 1.3875\times10^{-5}\left(\frac{T}{273.15}\right)^{\frac{3}{2}}\left(\frac{1}{P}\right),
\end{equation}
where $T$ is temperature and here $P$ is specifically in bar \citep{Chevrier:2008}, which gives nominal values on the order of 10$^{-4}$~m$^{2}$~s$^{-1}$ \citep{Schorghofer:2005, Chevrier:2007, Hudson:2007, Bryson:2008}. Fits to REMS derived water vapor pressure define the water vapor just above the regolith. Then, at the surface-atmosphere interface, a mass conservation boundary condition is applied, thereby coupling the REMS data to the model. Perturbative processes to simple diffusion, such as adsorption/desorption and frost formation are not included.  

Temperature was simulated to 4~m, beyond three times the annual skin-depth ($\sim$1~m), which allows for accurate modeling of temperature variations with depth and time, while mass transfer to a depth of 1~m. Simulations were run for several martian years and considered converged when the temperature with depth profile for two separate consecutive runs at the vernal equinox (Ls = 0$^{\circ}$) were $<1$~K different. We specifically tested the thermal inertia and albedo combinations for various terrains as found by \citet{Vasavada:2017}. Simulated surface temperatures and relative humidities were compared with surface derived REMS values. For each terrain traversed by MSL, the code had on average an error of $\pm5$~K in temperature and $\pm7\%$ in relative humidity with respect to REMS values and so was within error of MSL measurements. 

\subsection{Results}
In Figure 3, subsurface environmental conditions at varying depths are plotted in the calcium perchlorate phase diagram with respect to $RH_{l}$ for the low (3a), typical (3b), and maximum (3c) thermal inertia cases from \citet{Vasavada:2017}. Because simulations do not account for perturbative processes to water vapor diffusion, the maximum $RH_{l}$ is set to saturation with respect to ice. For low thermal inertia terrain, we find that deliquescence of calcium perchlorate is possible in the top few centimeters of the regolith between Ls 100$^{\circ}$ and 110$^{\circ}$ for up to one hour per sol for the case in Figure 3a ($\Gamma$ = 180~J~m$^{-2}$~K$^{-1}$~s$^{-1/2}$, A = 0.11). The rover was in this terrain unit between sols 1222 and 1242, or Ls $\sim94^{\circ}$ - $104^{\circ}$ \citep{Vasavada:2017}. For all higher thermal inertia terrains in Figure 3, brine formation is not favored in the subsurface.  

%
\begin{figure}[h!]
\centering
\includegraphics[width=20pc]{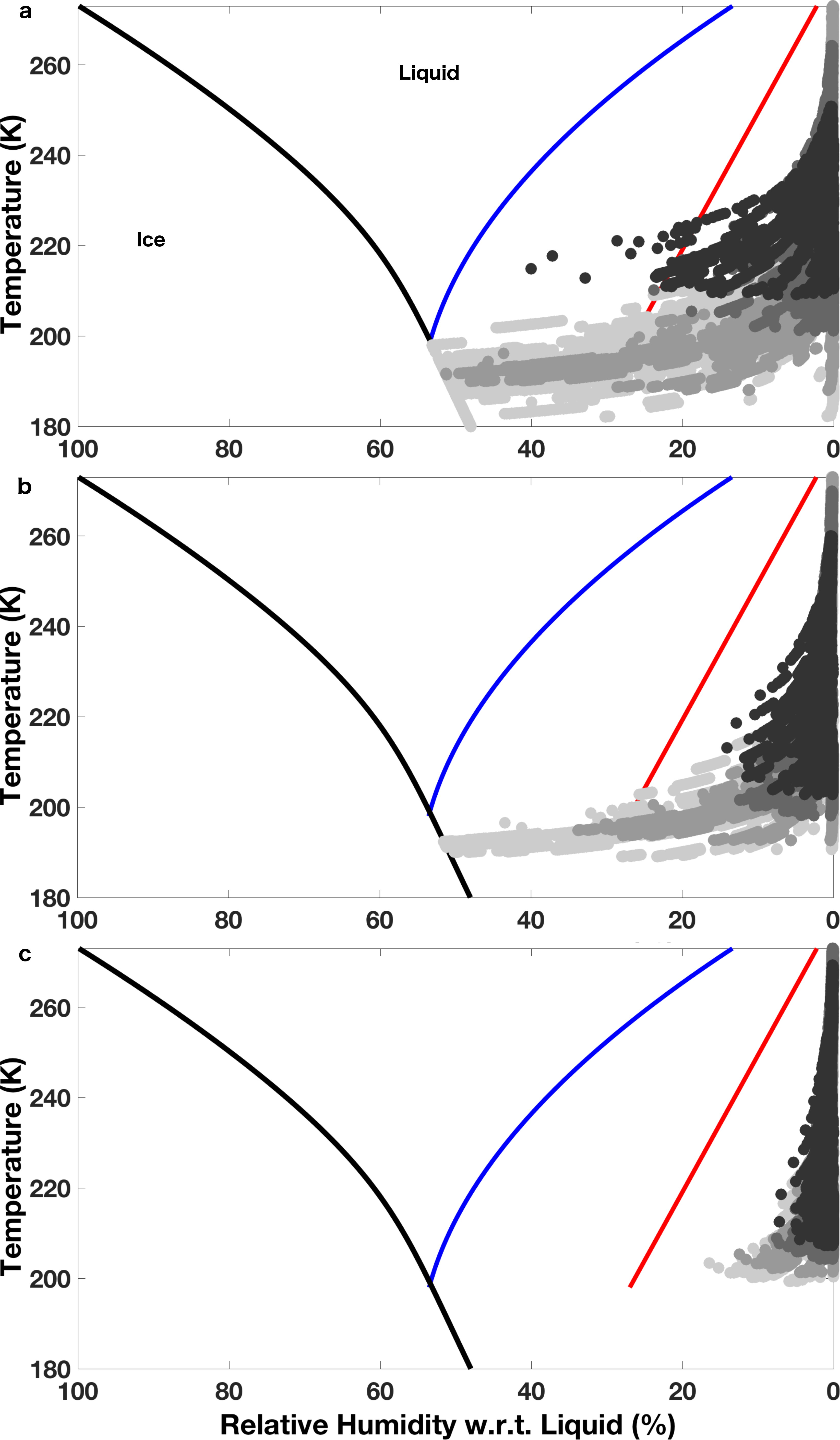}
\caption{Simulated subsurface conditions in temperature and $RH_{l}$ phase space for likely terrains throughout MSL's traverse, where (a) is $\Gamma$=180~J~m$^{-2}$~K$^{-1}$~s$^{-1/2}$, $A=0.11$, (b) $\Gamma=350$~J~m$^{-2}$~K$^{-1}$~s$^{-1/2}$, $A=0.2$, and (c) $\Gamma=550$~J~m$^{-2}$~K$^{-1}$~s$^{-1/2}$, $A=0.15$, as the low, typical, and high thermal inertia units traversed by the rover. Simulated hourly conditions for one martian year are in gray-scaled circles, where from lightest to darkest are results for depth of 1~cm, 2~cm, 4~cm, and 6~cm. Line colors follow from Figure 1.}
\end{figure}

To further analyze the potential for liquid formation at Gale crater, we tested combinations of thermal parameters as inferred in the first 1300 sols of the MSL mission \citep{Vasavada:2017}. Such simulations can inform on future MSL operation strategies and generally elucidate the potential for liquid formation through deliquescence in equatorial regions on Mars. Simulations were run from $150\leq\Gamma\leq300$~J~m$^{-2}$~K$^{-1}$~s$^{-1/2}$ in increments of $\Gamma=25$~J~m$^{-2}$~K$^{-1}$~s$^{-1/2}$, and for albedo from $0.1\leq A\leq0.3$ in increments of $A=0.05$. In Figure 4, we plot the thermal property parameter space along with the percent of the martian year brines are possible summed over the 1~m subsurface domain explored. Simulations suggest subsurface liquid formation through deliquescence of calcium perchlorate is not favored for terrains where $\Gamma>300$~J~m$^{-2}$~K$^{-1}$~s$^{-1/2}$. On the other hand, for $\Gamma\lesssim300$~J~m$^{-2}$~K$^{-1}$~s$^{-1/2}$ liquid formation is possible depending on the albedo while for $\Gamma\lesssim185$~J~m$^{-2}$~K$^{-1}$~s$^{-1/2}$ liquid formation is possible for a broad range of albedo values. Furthermore, results suggests conditions are the most apt for brine formation at low thermal inertia ($\Gamma\lesssim175$~J~m$^{-2}$~K$^{-1}$~s$^{-1/2}$) and high albedo ($A\gtrsim0.25$) terrains, where brines may be available for up to $\sim4\%$ of the year; however, such thermal inertia and albedo combinations have not been inferred at Gale crater by MSL through sol 1337 \citep{Vasavada:2017}. Most of the studied thermal inertia and albedo combinations inhibit subsurface brine formation, either entirely or limit it to a small fraction of the year. 

%
\begin{figure}[h!]
\centering
\includegraphics[width=30pc]{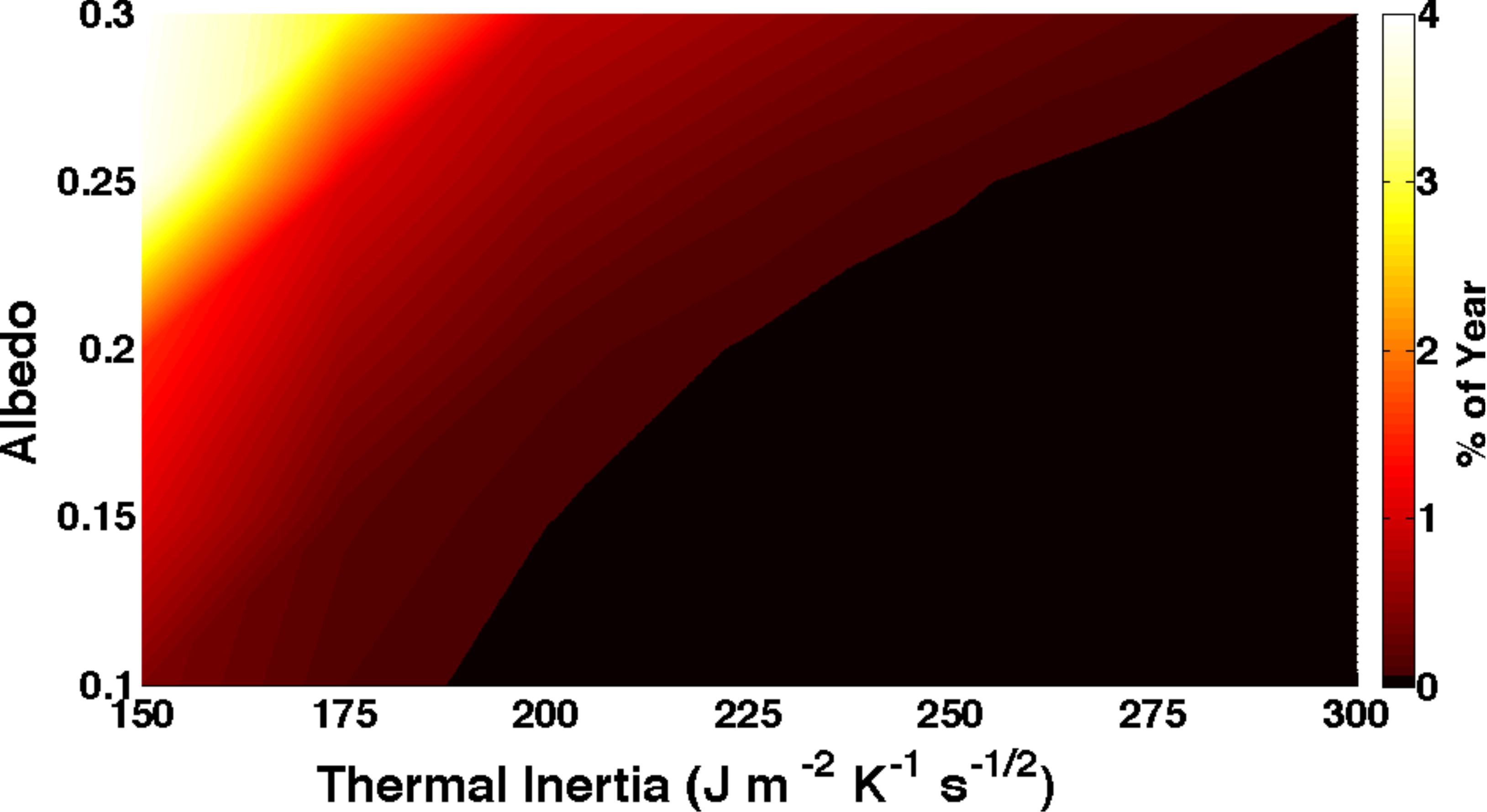}
\caption{The total percent of the year calcium perchlorate solutions may be possible within the subsurface (down to 1~m) at Gale crater as a function of thermal inertia and albedo. In the color map, 0\% is shown as black and values $>4$\% are white. Color map shading is interpolated between studied cases.}
\end{figure}

When brines are possible in the subsurface, the ambient conditions would permit a solution with a water activity of up to $a_{w}\sim0.55$, assuming equilibrium where $a_{w}=\left(RH_{l}/100\right)$. The temperature during the presence of brines is at most $T\sim205$~K. Although potential brines could meet the water activity criteria for Uncertain Regions \citep{Rummel:2014}, which requires $a_{w} \geq 0.5$, it does not simultaneously meet the temperature requirement of $T \geq 250$~K. Moreover, at these conditions the composition of the brine would be Ca(ClO$_{4}$)$_{2}$ = 52.4 wt\% and H$_{2}$O = 47.6 wt\%. Following reported perchlorate concentrations at Gale crater \citep{Leshin:2013} of 0.5~wt\% of Ca(ClO$_{4}$)$_{2}$, and assuming all the salt in the regolith were to deliquesce, the resulting solution would be $\sim0.5$~wt\% water, resulting in a liquid abundance in the regolith of $\sim1$~wt\% brine. We note this is an upper limit as it assumes all the salt deliquesces, and the atmosphere can provide sufficient water vapor. Therefore, even when liquids are potentially available in the subsurface, they are only present in small amounts.

%
%

\section{Discussion and Conclusions}
Liquid production through deliquescence has been suggested to be constrained to a select few regions on Mars \citep{Martinez:2013, Kossacki:2014}. Therefore, the potential signature of a fair amount of favorable conditions for liquid production at the surface of Gale crater \citep{MartinTorres:2015}, a near-equatorial location, could have implied a more global role for deliquescence. However, here we demonstrate that REMS derived surface environmental conditions in the first 1648 sols are not favorable to the formation of liquids through deliquescence of calcium perchlorate, which has the lowest known eutectic temperature for single component Mars-relevant brines. Accounting for error in the REMS data, this remains true at the 1-sigma level, but not at the 2-sigma level, where at two times the surface environmental conditions may have permitted liquid formation for up to an hour each day. These points occurred in active sand dunes on sols 1232 and 1311. Derived enthalpic changes may support liquid formation on sol 1232, but not necessarily sol 1311. Generally, though, derived enthalpic changes support the formation of frost at the surface of Gale crater as suggested by \citet{Martinez:2016}. As such, our results imply that surface liquid formation at Gale crater during MSL's traverse in the first 1648 sols was unlikely. 

Simulations of the subsurface environment at Gale crater for the terrains crossed by MSL, though, would suggest that low thermal inertia units could have been occasionally favorable to brine production through deliquescence in the shallow subsurface for a limited time between Ls 100$^{\circ}$ and 110$^{\circ}$. Therefore, simulations would support the formation of liquids on sol 1232, but in the shallow subsurface; however, they do not support liquid formation on sol 1311 as that is outside the predicted Ls range. Significant inferred enthalpic values (\emph{i.e.,} statistically non-zero), also typically occurred around Ls 100$^{\circ}$. Thus, these values may have indicated potential brine formation in the shallow subsurface as well.

A full study of the combination of thermal parameters (\emph{i.e.,} thermal inertia and albedo) suggests that brines may form in terrains with thermal inertia $\Gamma\lesssim300$~J~m$^{-2}$~K$^{-1}$~s$^{-1/2}$, depending on albedo, and may form for $\Gamma\lesssim185$~J~m$^{-2}$~K$^{-1}$~s$^{-1/2}$ for a broad range of albedo values. Subsurface brine formation, though, is most favorable in terrains with $\Gamma\lesssim175$~J~m$^{-2}$~K$^{-1}$~s$^{-1/2}$ and high albedo ($A\gtrsim0.25$). This could support the potential liquid involvement in martian slope streaks, which are found in dusty equatorial regions characterized by low thermal inertia \citep{Bhardwaj:2017}. The suggested combination of thermal properties required for significant production of subsurface liquids, though, has not yet been traversed by MSL through sol 1337 \citep{Vasavada:2017}. However, should the rover encounter such terrain in the future, our results could inform operational strategies for REMS and MSL's DAN (Dynamic Albedo of Neutrons) instrument, which is used to measure water-equivalent hydrogen in the subsurface \citep{Mitrofanov:2012}.

Potential subsurface brines would have water activities of up to $a_{w}\sim0.55$ and experience temperatures at most of $T\sim205$~K. As such, these liquids would not be considered habitable to life as we know it nor simultaneously meet the temperature and water activity requirements for Special or Uncertain Regions \citep{Rummel:2014}. Assuming typical perchlorate salt concentrations in the martian regolith and that all the salt enters solution, brine abundance in the regolith is also expected to be low, up to $\sim1$~wt\%, assuming the atmosphere can supply the water vapor. At these amounts, liquids formed by deliquescence of Ca(ClO$_{4}$)$_{2}$ may not support sediment transport, but could act as a trigger mechanism to instigate flow. In fact, recent experimental results suggest that grain levitation through vapor released by subsurface liquids may reduce the amount of fluid needed to form flow behavior by nearly an order of magnitude \citep{Raack:2017}. Furthermore, though recent models propose a granular flow mechanism for RSL formation, a trigger mechanism for flow initiation is still required, a potential role for liquids \citep{Dundas:2017}.

Our results suggest that only under a restricted set of conditions can calcium perchlorate solutions form at Gale crater, Mars; however, we only considered a single-salt brine system and simulated conditions on flat terrain. The martian regolith, though, is a mixture of various salts (\emph{e.g.,} \citet{Hecht:2009, Hanley:2012, ElSenousy:2015}). Brines with multiple dissolved salts may have lower eutectic temperatures and deliquescence relative humidities and thus be more stable under present-day martian conditions \citep{Gough:2014}. Therefore, once deliquescence is permitted in the subsurface of low thermal inertia terrains within Gale crater, dissolution may form multi-salt brines that would have increased stability relative to pure component solutions. Furthermore, slope effects on thermal insolation and horizontal distribution of water vapor will also affect the formation of brines. Indeed, as water vapor is transported across Gale crater, it is reduced due to surface-atmosphere interactions at the crater walls \citep{Steele:2017}. Consequently, the walls of Gale crater may provide more apt conditions to the formation of liquids through deliquescence. 

Simulations on the formation of subsurface brines at Gale crater suggest the search for liquids on present-day Mars by MSL may be most successful in the shallow subsurface of terrains characterized by low thermal inertia and high albedo, and specifically around Ls 100$^{\circ}$. Furthermore, using Gale crater as a proxy for equatorial regions on Mars, we suggest that humidity-temperature conditions are typically inconsistent with deliquescence in such regions. At most, the sum of the time the humidity-temperature conditions permit for deliquescence of calcium perchlorate in the subsurface (down to 1~m) is $4\%$. Terrains that may meet the required thermal properties ($\Gamma\lesssim185$~J~m$^{-2}$~K$^{-1}$~s$^{-1/2}$ for a broad range of albedos) to favor deliquescence account for only about half of the equatorial region on Mars \citep{Mellon:2000}. Even in these terrains, only small amounts of liquid production would be expected to form. Therefore, in equatorial regions, though deliquescence may play a role in triggering mass wasting events, such as potentially in RSL formation \citep{Dundas:2017}, it is unlikely to be a dominant atmosphere-regolith water vapor exchange process in contrast with polar regions, such as the Phoenix landing site, where deliquescence is expected to play a more active role in the near-surface exchange of water vapor \citep{Nuding:2014}.


%

\acknowledgments
This material is based upon work supported by the National Aeronautics and Space Administration (NASA) under Grant No. NNX15AM42G issued through the Mars Data Analysis Program. R. V. Gough acknowledges support from the NASA MSL Participating Scientist Program. K. M. Primm acknowledges support from NASA's Earth and Space Science Fellowship Grant No. NNX15AT60H. M. Tolbert acknowledges support from NASA's Mars Fundamental Research program through Grant No. NNX14AJ96G. G. M. Mart\'{i}nez acknowledges support from the Jet Propulsion Laboratory through Grant No. 1449038. All Mars Science Laboratory Rover Environmental Monitoring Station data used in this work are publicly available on the NASA Planetary Data System, Planetary Atmospheres Node. Data supporting Figures 3 and 4 are provided in the supplementary material. The authors thank the anonymous reviewers for valuable comments that helped improve this manuscript.

%




\listofchanges


\end{document}